# Patient-Specific Cerebral Aneurysm Hemodynamics: Comparison of *in vitro* Volumetric Particle Velocimetry, Computational Fluid Dynamics (CFD), and *in vivo* 4D Flow MRI

Melissa C. Brindise[1], Sean Rothenberger[2], Benjamin Dickerhoff[3], Susanne Schnell[4], Michael Markl[4,5], David Saloner[6], Vitaliy L. Rayz[2], Pavlos P. Vlachos[1]

[1]*School of Mechanical Engineering, Purdue University, West Lafayette, IN*

[2]*Weldon School of Biomedical Engineering, Purdue University, West Lafayette, IN*

[3]*Department of Biomedical Engineering, Marquette University, Milwaukee, WI*

[4]*Feinberg School of Medicine, Northwestern University, Chicago, IL*

[5]*McCormick School of Engineering, Northwestern University, Evanston, IL*

[6]*Department of Radiology and Biomedical Imaging, University of California San Francisco, CA*

Corresponding Author:

Pavlos P. Vlachos

Email: pvlachos@purdue.edu

Address: 585 Purdue Mall, West Lafayette, IN, 47907

## Abstract

Typical approaches to patient-specific hemodynamic studies of cerebral aneurysms use image based computational fluid dynamics (CFD) and seek to statistically correlate parameters such as wall shear stress (WSS) and oscillatory shear index (OSI) to risk of growth and rupture. However, such studies have reported contradictory results, emphasizing the need for in-depth comparisons of volumetric experiments and CFD. In this work, we conducted tomographic particle velocimetry experiments using two patient-specific cerebral aneurysm models (basilar tip and internal carotid artery) under pulsatile flow conditions and processed the particle images using Shake the Box (STB), a particle tracking method. The STB data was compared to that obtained from *in vivo* 4D flow MRI and CFD. Although qualitative agreement of flow pathlines across modalities was observed, each modality maintained notably unique spatiotemporal distributions of low normalized WSS regions. Analysis of time averaged WSS (TAWSS), OSI, and Relative Residence Time (RRT) demonstrated that non-dimensional parameters, such as OSI, may be more robust to the varying assumptions, limitations, and spatial resolutions of each subject and modality. These results suggest a need for further multi-modality analysis as well as development of non-dimensional hemodynamic parameters and correlation of such metrics to aneurysm risk of growth and rupture.

# 1 Introduction

It is estimated that about 3% of the population harbors an unruptured intracranial aneurysm (IA) [1]. If detected, clinicians must assess and balance risk of rupture with risk of treatment of the cerebral aneurysm [2,3]. However, accurately assessing risk of rupture in IAs is difficult as the specific mechanisms that cause an aneurysm to form, grow, and rupture remain largely unknown.

Previous studies have demonstrated that hemodynamics play a critical role in the growth and rupture of an IA [4–8]. However, despite a large volume of studies that have investigated the influence of several hemodynamic variables on risk of rupture, contradictory and ultimately inconclusive results have been reported. Wall shear stress (WSS) has received much attention, but has been perhaps the most controversial hemodynamic parameter [8]. Both low [4,9–12] and high [13,14] WSS have been shown to indicate elevated risk of rupture. High WSS gradients, high oscillatory shear index (OSI), and high relative residence time (RRT) have also been reported to increase risk of rupture [4,8,11,15]. Studies have also explored the presence and effect of chaotic flow and high frequency fluctuations in cerebral aneurysms [3,6,16]. Other hemodynamic variables identified as potentially increasing risk of rupture include concentrated inflow jets, larger shear concentration, lower viscous dissipation, and complex and unstable flow patterns [6,13,14].

Computational fluid dynamics (CFD) has been the predominant methodology used to study hemodynamics in cerebral aneurysms [2,4,9,10,14,16]. However, limited CFD validation and disputed CFD assumptions such as laminar flow remain issues limiting its clinical acceptance [17]. It has also been shown that CFD results can vary significantly based on solver parameters, even when similar geometries and boundary conditions are used [5]. Velocity fields obtained from both *in vivo* and *in vitro* 4D Flow MRI have also been used, but the low spatiotemporal resolution is a major limitation of this modality [5,10,18]. Although high resolution 4D Flow MRI has been shown to capture complex flow patterns [18], studies have demonstrated that the low resolution causes an underestimation of velocity and WSS magnitudes, particularly when compared to CFD or PIV [2,5,10,19]. Few experimental particle image velocimetry (PIV) studies have been conducted in this domain. Among such studies, planar and stereo PIV (2D-2 velocity component) have been most common [3,11,15,18,20]. Other studies have taken 2D data at several parallel planes in a “sliced planar” (3D-2 velocity component) configuration and subsequently reconstructed the third velocity component [6,21]. Despite the high three-dimensionality known to exist in cerebral aneurysm geometries, to date, only one steady flow 4D volumetric PIV (3D-3 velocity component) study has been reported [5]. Further, PIV studies are most often used only to compare flow patterns and validate CFD models and simulations [5,11,18], only a few have reported WSS and even less OSI and RRT [6,15].

In this work, we conducted one of the first reported pulsatile volumetric particle velocimetry studies using two patient-specific aneurysm models. As opposed to traditional PIV studies which use interrogation window correlation, we used Shake the Box (STB), a particle tracking methodology, which has, to the authors’ knowledge, never been used in this domain. We compare the STB results with those obtained using both CFD and *in vivo* 4D Flow MRI. All three modalities maintain different assumptions and limitations that can affect subsequent hemodynamic analysis. In this work, we investigate the effect of spatial resolution and modality on time averaged WSS (TAWSS), OSI, and RRT.

# 2 Materials and Methods

## 2.1 *In vivo* 4D Flow MRI and MRA imaging

**Table 1: 4D Flow MRI parameters and resolutions**

| Geometry | TE/TR (ms) | Flip Angle (°) | Velocity Encoding (venc) (cm/s) | Temporal Resolution (ms) | Spatial Resolution (mm) |
|---|---|---|---|---|---|
| Basilar Tip | 3.46/6.33 | 12 | 100 | 40.5 | 1.25 x 1.25 x 1.33 |
| ICA | 2.997/6.4 | 15 | 80 | 44.8 | 1.09 x 1.09 x 1.30 |

The two aneurysm models used were a basilar tip aneurysm, MRI imaged at the San Francisco VA Medical Center, and an internal carotid artery (ICA) aneurysm, MRI imaged at Northwestern Memorial Hospital (NMH). Both aneurysms were imaged on a 3T MRI scanner (Skyra, Siemens Healthcare, Erlangen, Germany). At San Francisco VA Medical Center, the Siemens WIP sequence, an ECG-gated RF spoiled 4D Flow MRI sequence was used with gadolinium contrast, while at NMH no contrast was used. The 4D flow scan parameters for both imaging studies are given in Table 1. The *in vivo* 4D Flow MRI data will be referred to simply as '4D Flow' herein.

All 4D Flow data was corrected for noise, velocity aliasing, and phase offset errors caused by eddy currents and concomitant gradient terms. In addition to 4D Flow, contrast-enhanced magnetic resonance angiography (CE-MRA) data with a spatial resolution of 0.7 x 0.7 x 0.7 mm$^3$ for the basilar tip aneurysm and non-contrast time of flight (TOF) angiography with a spatial resolution of 0.4 x 0.4 x 0.6 mm$^3$ for the ICA aneurysm were acquired in the same scanning sessions and used to create the *in vitro* models.

## 2.2 Image segmentation and model fabrication

For the basilar tip aneurysm, CE-MRA images were segmented using an in-house VTK-based software. A 3D iso-surface was computed by selecting a threshold intensity value that defined the intra-luminal volume of the vessel. The threshold was adjusted to match the iso-surface to the MR luminal boundaries. For the ICA aneurysm, TOF images were segmented with open-source ITK-SNAP software, using thresholding and volume growth techniques. A modeling software, Geomagic Design (3D Systems, Rock Hill, SC), was used to separate the inflow and outflow vessels of the aneurysm from the remaining cerebral vasculature (Figures 1a and 1c). The resulting surface (STL), shown in Figures 1b and 1d, was used for CFD simulations and flow phantom fabrication. For the flow phantoms, inlet and outlet vessels were extended in order to connect the model to the flow loop. A positive-space model of the vascular geometry was 3D printed (ProJet printer - 3D Systems), embedded into a tear-resistant silicone block, then cut from the block. A low melting point metal (Cerrobend 158 Bismuth alloy) was cast into the block and the block was then cut away. The metal model was embedded in optically clear polydimethylsiloxane silicone (PDMS—Slygard 184) which was allowed to cure until hardened, then the metal was melted out from the clear PDMS.

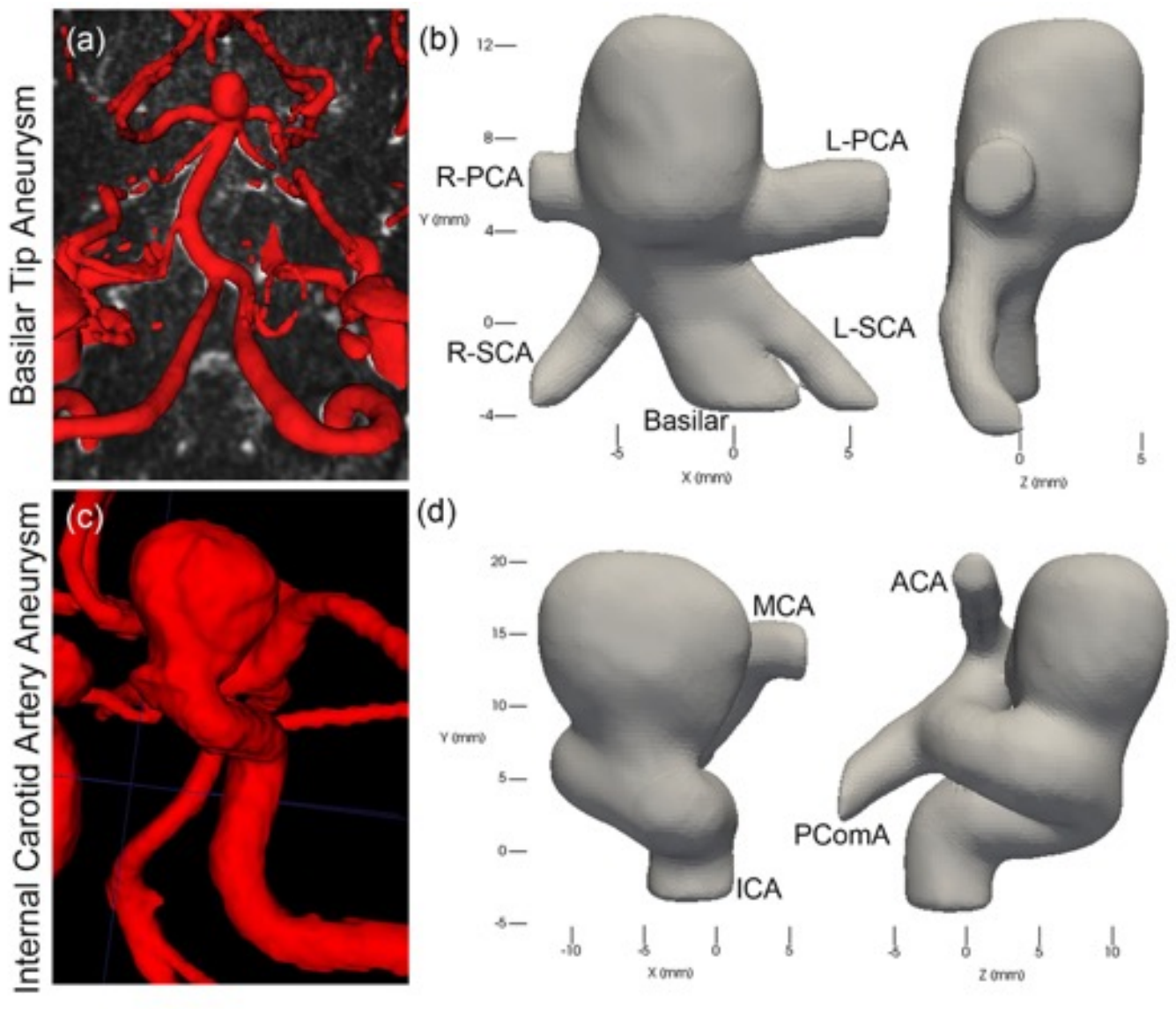

**Figure 1: Segmentation from (a) *in vivo* vasculature showing patient-specific basilar tip aneurysm, to the (b) associated *in vitro* model. (c) *In vivo* vasculature for patient-specific internal carotid artery (ICA) aneurysm and the (d) *in vitro* segmented model. (PCA=posterior cerebral artery, SCA=superior cerebellar artery, MCA=middle cerebral artery, ACA=anterior cerebral artery, PComA = posterior communicating artery).**

### 2.3 *In vitro* flow loop

An *in vitro* flow loop (Figure 2a) was designed to simulate *in vivo* flow conditions. The working fluids, detailed in Table 2, were blood analogs consisting of water, glycerol, and urea [22]. The pulsatile inflow waveforms (Figure 2b) were extracted from the *in vivo* 4D Flow MRI data and generated by a computer-controlled gear pump. Average flow rates for the outlet vessels were controlled using resistive elements to match the 4D Flow outlet flow rate ratios. Flow details for both geometries are provided in Table 3.

**Table 2: Blood analog working fluids used for both geometries. Nano-pure water, technical grade glycerol (99% - McMaster-Carr), and 99+% urea (Fischer Scientific) were used.**

| Geometry | Blood Analog Composition (%wt) Water | Glycerol | Urea | Density (kg/m$^3$) | Kinematic Viscosity (m$^2$/s) |
|---|---|---|---|---|---|
| Basilar Tip | 44.8 | 32.8 | 22.4 | 1103 | 3.04x10$^{-6}$ |
| ICA | 45.3 | 29.7 | 25.0 | 1132 | 3.50x10$^{-6}$ |

**Table 3: Flow cycle information for both aneurysm geometries**

| Geometry | Reynolds Number Max | Min | Womersley Number | Cycle Period, T (sec) |
|---|---|---|---|---|
| Basilar Tip | 500 | 150 | 2.73 | 0.77 |
| ICA | 300 | 130 | 5.33 | 0.54 |

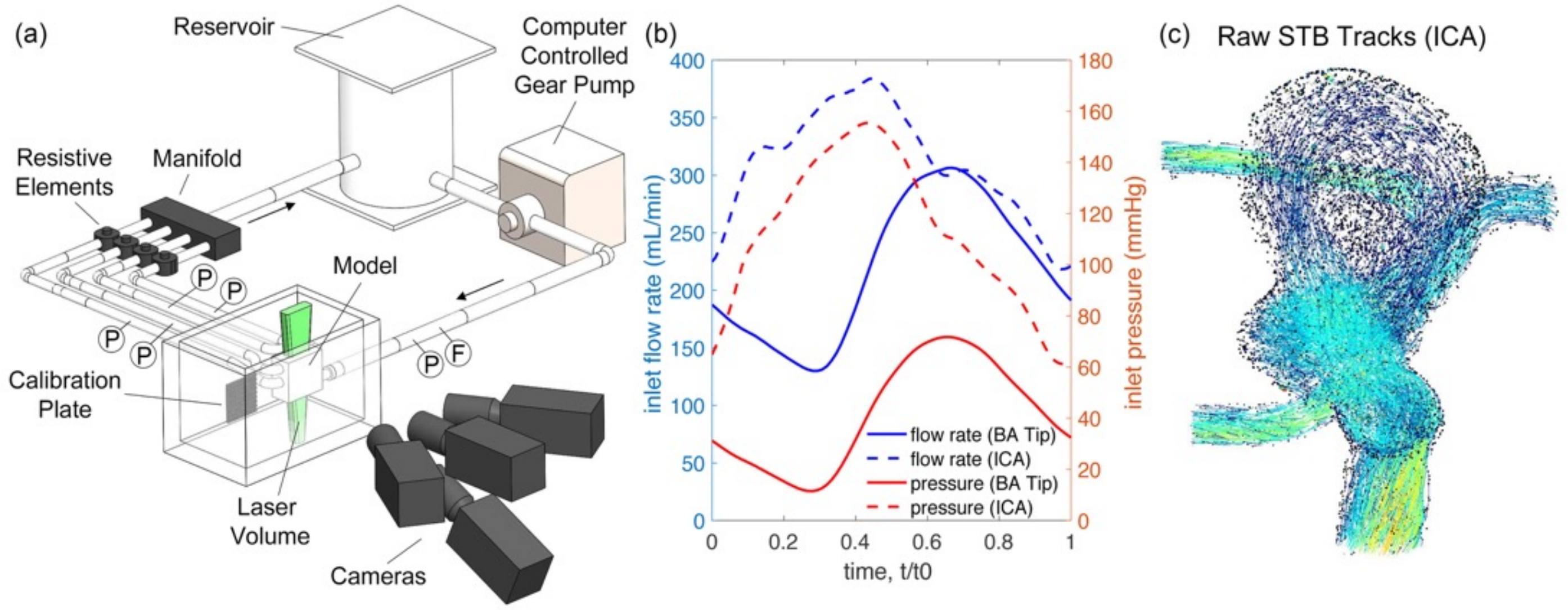

**Figure 2: (a) Schematic of the flow loop setup, including the camera and calibration plate setup. F indicates locations of ultrasonic flowmeters and P indicates locations of pressure transducers. (b) Inflow flow rate and pressure taken from the upstream flow meter and pressure transducers for both geometries. The phase of the pulsatile cycle is displayed as it was extracted from 4D Flow MRI. (c) Sample Shake the Box tracks for the ICA geometry.**

### 2.4 Volumetric particle velocimetry measurement technique

Particle images were captured using an Nd-YLF laser (Continuum Terra-PIV, $\lambda$ = 527 nm) and four high-speed cameras (Phantom Miro). In the camera configuration (Figure 2a), the center camera was not angled, while the other three were angled about 30° from the geometry plane. The magnification of all cameras was approximately 30 μm/pixel. Flow was seeded with 24 μm fluorescent particles and long-pass filters filtered the laser light from the images. The index of refraction of the working fluid was matched to that of the PDMS model (n = 1.4118), and the model was submerged in the working fluid to reduce optical distortion. Time-resolved images (1216x1224 pixels for basilar tip, 1088x1320 pixels for ICA) were captured at 2000 Hz, corresponding to a maximum particle displacement between frames of approximately 10 pixels. Three full pulsatile cycles for the basilar tip aneurysm and four cycles for the ICA aneurysm were captured.

Particle images were processed using DaVis 10.0 (LaVision Inc.). Calibration images were captured using a dual-plane calibration target. The perspective calibration error was approximately 0.25 pix for each camera in both experiments and volume self-calibration corrected these errors to less than 0.03 pix [23]. Rather than using 3D volume correlation as used in tomographic PIV, Shake the Box (STB), a particle tracking based algorithm, was used to process the velocity fields. Sample STB tracks are shown in Figure 2c. STB reduced the prevalence of ghost particles (falsely triangulated 3D particles), which are expected to be high in these complex geometries and can create significant bias error [24,25]. STB requires time-resolved data, using the previous time step to iteratively add predicted particle locations in the next time step. The predicted particle locations are refined by iteratively "shaking" them within a tolerance, minimizing the residual between the subsequent particle image and predicted particle image. The STB was done using 12 iterations for both the outer (adding particles) and inner (particle position refinement) loops with an allowed particle triangulation error of 1.5 voxels and particle position shaking of 0.1 voxels. Currently, no methods to evaluate uncertainty using STB processing have been reported, however, the calibration errors were consistent with those observed in well-controlled tomographic PIV experiments. The volumetric STB data will be referred to as 'STB' herein.

### 2.5 CFD simulation

CFD simulations were performed using Fluent 18.1 (ANSYS) to solve the governing Navier-Stokes equations. An unstructured tetrahedral mesh was generated on the domain using HyperMesh 14.0 (Altair Engineering, Troy, MI). Incompressible, Newtonian flow with density of 1060 kg/m$^3$ and dynamic viscosity of 0.0035 Pa·s was modeled. Laminar flow and rigid walls were assumed. Patient-specific waveforms obtained from 4D Flow were prescribed at the inlet and outlets of the CFD models. A pressure-based coupled algorithm was used to solve the momentum and pressure-based continuity equations. A second-order Crank–Nicolson scheme was used in time discretization and a third-order MUSCL scheme was used for discretization of the momentum equations. A mesh with the nominal cell size of 150 μm and time step of 1.5 ms was used. These values provided sufficient resolution of the flow based on mesh independence and temporal resolution testing. Three cardiac cycles were simulated, and the results obtained for the last cycle were used for comparisons with the other modalities.

### 2.6 Post-processing

All modalities were registered to and masked by the STL geometry obtained from MRA segmentation. The unstructured STB and CFD data were gridded to isotropic resolutions of 0.3 and 0.4 mm for the basilar tip and ICA aneurysms, respectively, using inverse-squared radial distance weighted averaging. An initial radius of half the grid size was used to search for the nearest unstructured vectors, however this radius was extended to ensure a minimum of three unstructured vectors for each averaging calculation. For the STB gridding, unstructured velocity fields were temporally grouped at a 5:1 and 3:1 ratio for the basilar tip and ICA aneurysms, respectively, such that each unstructured field was used in only a single group. This increased the number of unstructured particles per gridded time step but reduced the effective temporal resolution to 2.5 ms and 1.5 ms for the basilar tip and ICA aneurysms, respectively. The gridded STB data was filtered using proper orthogonal decomposition (POD) with the entropy line fit (ELF) autonomous thresholding method [26,27]. A single pass of universal outlier detection (UOD) [28] and phase averaging of the STB pulsatile cycles were subsequently done. “Virtual voxel averaging” was spatially performed to bring the STB and CFD data to the 4D Flow spatial resolution.

Wall shear stress was computed using thin-plate spline radial basis functions (TPS-RBF), which performs smoothing surface fits and reduces errors in the inherently noisy gradient calculation [29]. A normal vector at each surface point was computed by mapping the velocity coordinate to the equivalent STL-surface point and computing the inward facing normal from a surface fit of 25 adjacent STL-surface points. Two passes of UOD were used to eliminate erroneous normal vectors. For the voxel averaged datasets, wall normals were extracted from their full resolution equivalent. For the 4D Flow, the full resolution CFD was used. WSS was computed according to the following equations [30]:

$$\tau_x = \mu\left(2n_x\left(\frac{du}{dx}\right) + n_y\left(\frac{du}{dy} + \frac{dv}{dx}\right) + n_z\left(\frac{du}{dz} + \frac{dw}{dx}\right)\right) \quad (2.1)$$

$$\tau_y = \mu\left(n_x\left(\frac{du}{dy} + \frac{dv}{dx}\right) + 2n_y\left(\frac{du}{dy}\right) + n_z\left(\frac{dv}{dz} + \frac{dw}{dy}\right)\right) \quad (2.2)$$

$$\tau_z = \mu\left(n_x\left(\frac{du}{dz} + \frac{dw}{dx}\right) + n_y\left(\frac{dv}{dz} + \frac{dw}{dy}\right) + 2n_z\left(\frac{dw}{dz}\right)\right) \quad (2.3)$$

$$\tau_{mag} = \sqrt{\tau_x^2 + \tau_y^2 + \tau_z^2} \quad (2.4)$$

where $\tau_x$, $\tau_y$, and $\tau_z$, are the WSS components in the x, y, and z directions, $\tau_{mag}$ is the WSS magnitude, $\mu$ is the dynamic viscosity, and $(n_x, n_y, n_z)$ is the unit normal.

The in-house WSS code, including the wall normal and velocity gradient calculations, was validated using analytical 3D Poiseuille flow data. Biases in the near-wall discrete velocity gradients, whose magnitude vary based on each specific point’s distance from the wall, is a known issue [6,29]. Thus, similar to that done in Yagi et al. (2013) [6], the WSS calculation was extended to use gradients beyond the biased near-wall region, mitigating the spatial variation of this bias to about 5% but allowing WSS magnitude bias errors of up to 25% (based on the validation testing). Thus, all WSS magnitudes reported here are expected to have a consistent bias and should be considered only relative to other values reported here.

Time averaged WSS (TAWSS) was computed according to Eq. 2.5

$$TAWSS = \frac{1}{T}\int_0^T |\tau_w| dt \quad (2.5)$$

where $\tau_w$ is the WSS vector and T is the duration of the pulsatile cycle. Oscillatory shear index (OSI) was subsequently computed according to Eq. 2.6

$$OSI = \frac{1}{2}\left(1 - \frac{\frac{1}{T}\left|\int_0^T \tau_w dt\right|}{\frac{1}{T}\int_0^T |\tau_w| dt}\right). \quad (2.6)$$

OSI is a non-dimensional parameter ranging from 0 to 0.5, where 0 indicates no oscillatory WSS throughout the pulsatile cycle and 0.5 indicates oscillatory WSS throughout the entire pulsatile cycle. Relative residence time (RRT) was computed using Eq. 2.7

$$RRT = \frac{1}{(1-2OSI)TAWSS} \ . \quad (2.7)$$

# 3 Results

## 3.1 Comparing flow structures and WSS across modalities

The inlet flow rate computed from the velocity fields for all modalities in the basilar tip and ICA aneurysms were computed to ensure agreement across modalities and are shown in Figures 3a and 3b, respectively. Because of the low spatial resolution of 4D Flow, a larger uncertainty of the computed flow rate was expected. The maximum inlet flow rate in the basilar tip aneurysm was 4.28, 4.89, and 5.61 mL/s for the 4D Flow and full resolution STB and CFD, respectively. The average inlet flow rate was 2.58 mL/s for the 4D Flow, 3.48 mL/s for STB, and 3.51 mL/s for CFD. The trend of the inflow waveforms was similar for all modalities, with STB maintaining the largest temporal variability among modalities as expected. For the ICA aneurysm, the maximum inflow rate was similar for all modalities at 5.46, 5.40, and 5.26 mL/s for the 4D Flow, and full resolution STB and CFD, respectively. The average inlet flow rate was 4.26 mL/s for 4D Flow, 4.00 mL/s for STB, and 3.79 mL/s for CFD. Again, good agreement of the inflow waveform trend was observed across all modalities.

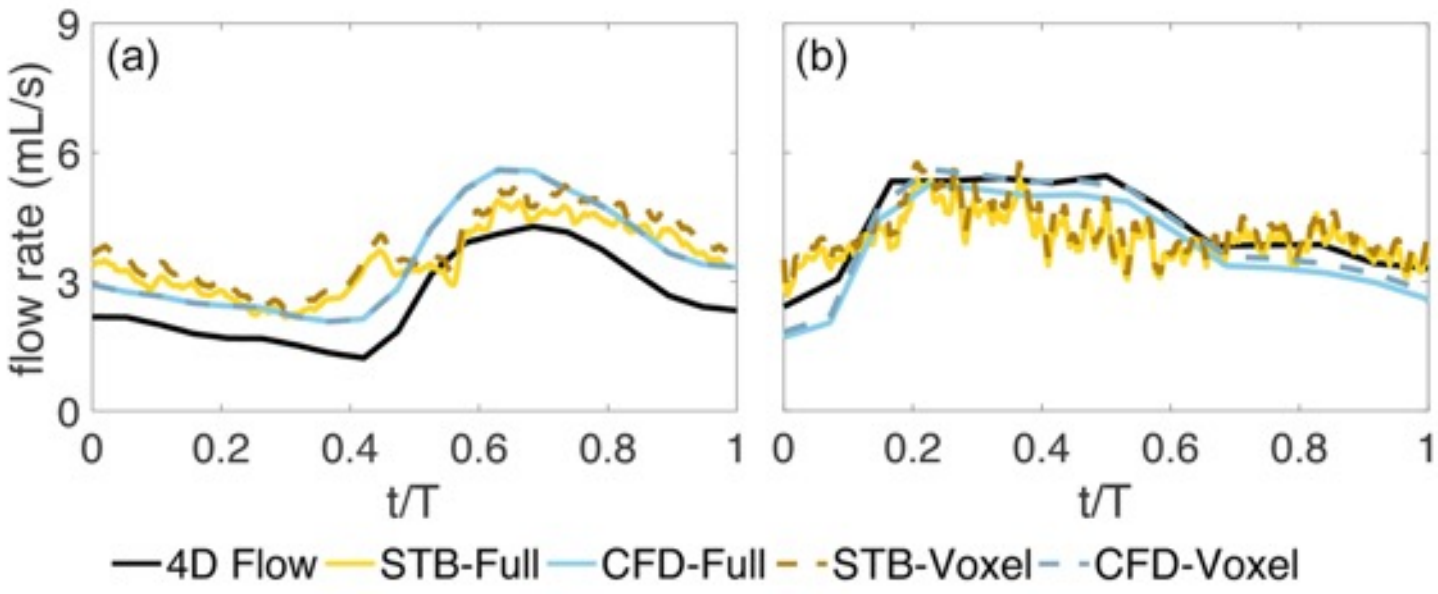

**Figure 3: Inlet flow rates for all modalities in the (a) basilar tip aneurysm and (b) ICA aneurysm.3**

Figure 4 shows the 3D velocity pathlines for each modality and geometry at peak systole. Initial observation shows qualitative agreement of the flow patterns across all modalities for both geometries. In the basilar tip aneurysm (Figure 4a) flow enters from the basilar artery and about 20% exits through the superior cerebellar arteries (SCAs). The remainder of the flow extends into the right distal posterior region of the aneurysmal sac, swirls to the left proximal anterior portion of the sac and exits the posterior cerebral arteries (PCAs). In the ICA aneurysm (Figure 4b) flow enters from the curving ICA and some flow rotates through the aneurysmal sac before re-entering the distal ICA. About half of the flow exits through the middle cerebral artery (MCA) and the remainder splits between the anterior cerebral artery (ACA) and posterior communicating artery (PComA). The swirling flow in the aneurysmal sac was qualitatively similar for the 4D Flow and CFD, while for the STB the swirling was more centered in the sac.

Figure 5 illustrates the WSS distribution at peak systole, normalized by the maximum peak systole WSS for each modality. The same WSS calculation method was used for all modalities. For both geometries, despite the similarities in the flow rate and flow structure, significant spatial differences among the WSS across all modalities were observed. For the basilar tip aneurysm (Figure 5a), the full resolution CFD exhibited a large region of low WSS at the proximal anterior portion of the aneurysmal sac. Conversely, the full resolution STB showed no region of low WSS and 4D Flow maintained a much smaller region of low WSS at that location. For the ICA aneurysm (Figure 5b), 4D Flow, STB, and CFD all indicated uniquely different regions of low WSS. In this case, STB indicated the entire aneurysmal sac having low WSS, while CFD results showed patches of low WSS at the anterior and posterior faces of the aneurysmal sac, and 4D Flow showed no significant low WSS regions. Except for the STB results in ICA aneurysm, no regions of low WSS among the voxel averaged datasets were observed. In terms of WSS magnitude, for the basilar tip aneurysm (Figure 5c), the full resolution STB and CFD had similar PDF distributions, with STB maintaining slightly higher magnitudes. 4D Flow had slightly lower WSS magnitude than that obtained in STB and CFD. The voxel averaged STB and CFD maintained similar distributions of WSS magnitude. For the ICA aneurysm (Figure 5d), the WSS magnitude distributions of the 4D Flow and full resolution STB matched well. The voxel averaged STB and CFD distributions also had similar distributions to that of the 4D Flow, with slight under and over estimations comparatively. Conversely, the full resolution CFD WSS distribution maintained a larger spread of WSS values than all other modalities and maintained the largest WSS magnitudes.

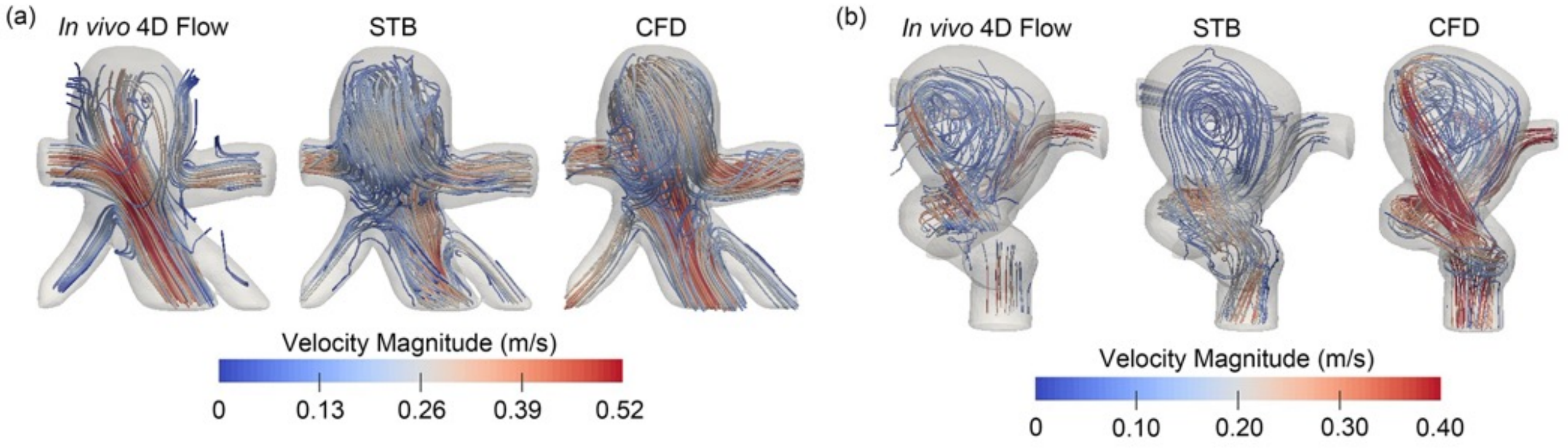

**Figure 4: Velocity field pathlines for the MRI and full resolution STB and CFD at peak systole for the (a) basilar tip aneurysm and (b) ICA aneurysm. (Note: The two aneurysm geometries (a) and (b) are not shown at the same spatial scale.)**

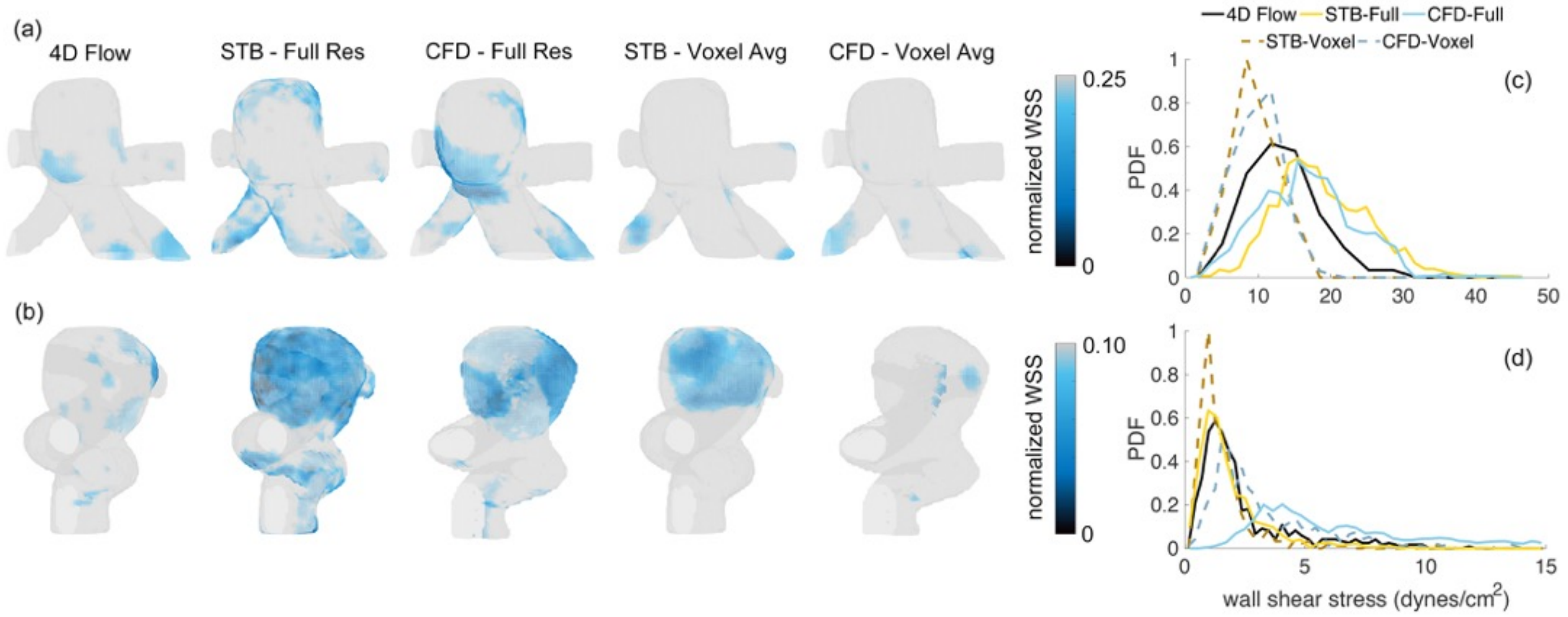

**Figure 5: Normalized wall shear stress distribution for all modalities in the (a) basilar tip aneurysm and (b) ICA aneurysm. (Note: The two aneurysm geometries (a) and (b) are not shown at the same spatial scale.) The PDF of wall shear stress in the aneurysmal sac for all modalities and resolutions in the (c) basilar tip aneurysm and (d) ICA aneurysm.**

### 3.2 Effect of modality and spatial resolution on TAWSS, OSI, and RRT

The previous section's results suggested that WSS is highly sensitive to the methods, assumptions and limitations associated with different modalities, even when flow rates and flow structures showed good agreement. To further explore this notion, the influence of the different modalities and effect of spatial resolution on TAWSS, OSI, and RRT was evaluated.

Figure 6 shows the distributions of TAWSS, OSI, and RRT in the aneurysmal sac for all modalities for the basilar tip aneurysm (Fig 6a-c) and ICA aneurysm (Fig 6d-f). The interquartile ranges and 95% confidence intervals are also provided. In agreement with the previous results, the distributions of the TAWSS varied significantly across all five data sets. In the basilar tip aneurysm, the average TAWSS was 9.21 dynes/cm$^2$ for the 4D Flow, 16.14 dynes/cm$^2$ for the full resolution STB, and 11.61 dynes/cm$^2$ for the full resolution CFD. When voxel averaged, the average TAWSS decreased to 7.39 dynes/cm$^2$ for the STB, a 54% drop, and to 6.82 dynes/cm$^2$ for the CFD, a 41% decrease. For the ICA aneurysm, the average TAWSS for STB had only a small change from 2.28 to 2.06 dynes/cm$^2$ when voxel averaged. The average CFD TAWSS decreased 70% from 8.25 to 2.46 dynes/cm$^2$ when voxel averaged. The average OSI of the STB for the basilar tip aneurysm changed from 0.12 to 0.11, while CFD went from 0.04 to 0.07 when voxel averaged. Voxel averaging changed the average OSI in the ICA aneurysm from 0.16 to 0.09 for STB and from 0.06 to 0.08 for CFD. In terms of RRT, the average RRT was 0.47, 0.28, and 0.29 (dynes/cm$^2$)$^{-1}$ for the 4D Flow, and full resolution STB and CFD, respectively, in the basilar tip aneurysm. When voxel averaged, these values changed to 1.22 (dynes/cm$^2$)$^{-1}$ for STB and 0.50 (dynes/cm$^2$)$^{-1}$ for CFD. Average RRT in the ICA aneurysm was 1.93, 1.38, and 0.42 (dynes/cm$^2$)$^{-1}$ for the 4D Flow, and full resolution STB and CFD, respectively. Voxel averaging increased the average RRT to 2.62 (dynes/cm$^2$)$^{-1}$ for STB and to 1.27 (dynes/cm$^2$)$^{-1}$ for CFD. Overall, the median OSI change was about 40% when data was voxel averaged while the median TAWSS and RRT changes were 48 and 145%, respectively.

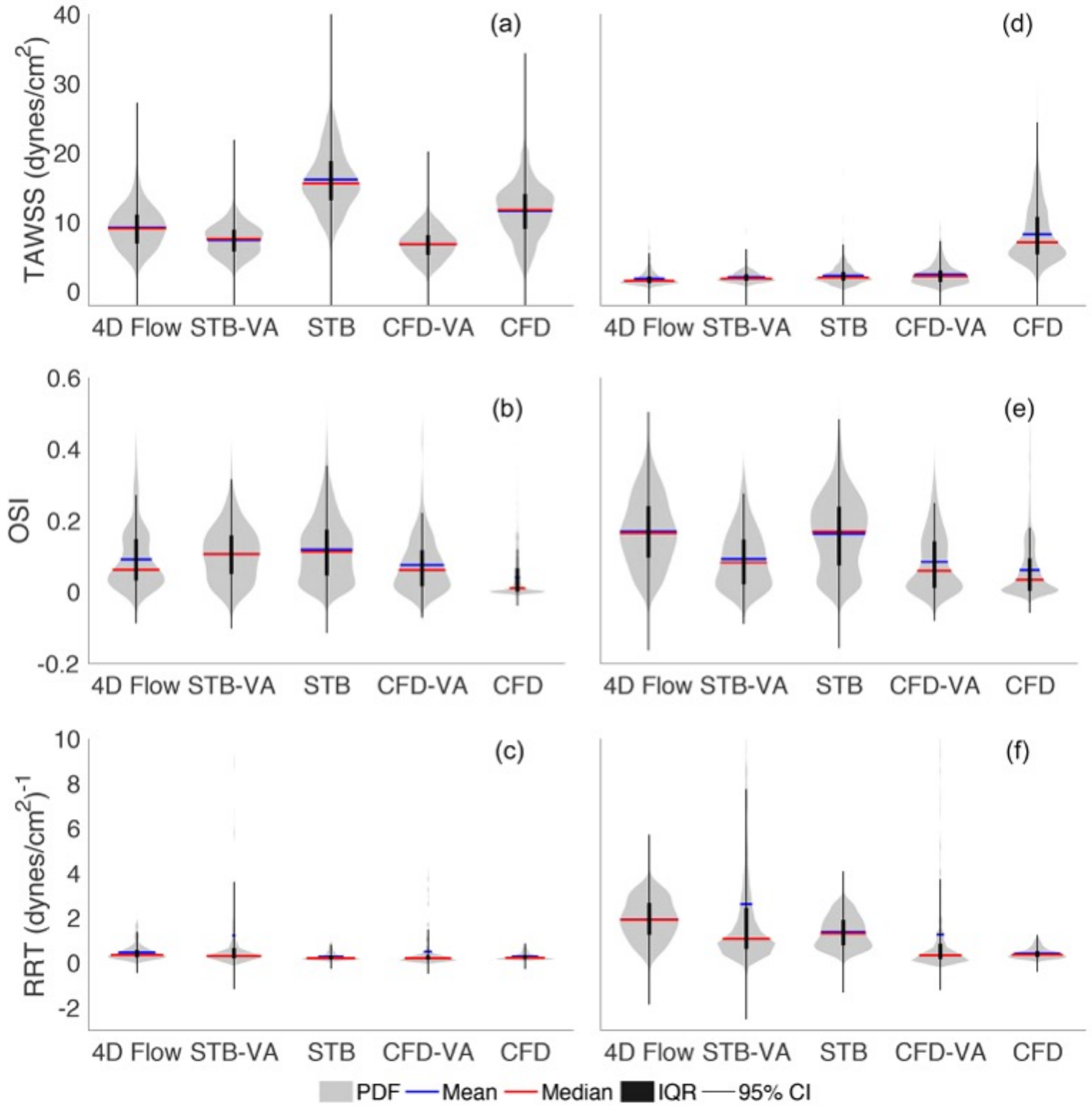

**Figure 6: Distribution of (a) time averaged wall shear stress (TAWSS), (b) oscillatory shear index (OSI), and (c) relative residence time (RRT) in the basilar tip aneurysm, where width of PDF shows relative distribution density. Distribution of (d) TAWSS, (e) OSI, and (f) RRT in the ICA aneurysm. Mean and median values, interquartile ranges, and 95% confidence intervals are indicated.**

To investigate the effect of spatial resolution on TAWSS, OSI, and RRT in more detail, Figure 7 illustrates Bland-Altman analysis comparing the voxel averaged and full resolution STB and CFD across the entire flow domain. In both aneurysms, a proportional error of the TAWSS was observed where the voxel averaged TAWSS magnitude was less than that of the full resolution, in agreement with the results in Figure 6. The average differences were -10.52 and -6.02 dynes/cm$^2$ for the STB and CFD, respectively, in the basilar tip aneurysm (Fig 7a). For the ICA aneurysm (Fig 7b), the average TAWSS difference was -0.16 dynes/cm$^2$ for STB and -8.70 dynes/cm$^2$ for CFD. The average OSI difference magnitude across all cases was 0.02, with a maximum difference magnitude of 0.05. The spread of OSI points was relatively symmetric, demonstrating no significant bias or proportional error. In the basilar tip aneurysm, the average RRT difference was 0.77 (dynes/cm$^2$)$^{-1}$ for the STB and 0.35 (dynes/cm$^2$)$^{-1}$ for the CFD, and 2.27 and 0.51 (dynes/cm$^2$)$^{-1}$ for the STB and CFD, respectively, in the ICA aneurysm. A similar proportional type error was observed with the RRT as with the TAWSS, but in this case the voxel averaged RRT was larger in magnitude than that of the full resolution. Thus, Figure 7 confirms the varying behavior and sensitivity of each metric to spatial resolution.

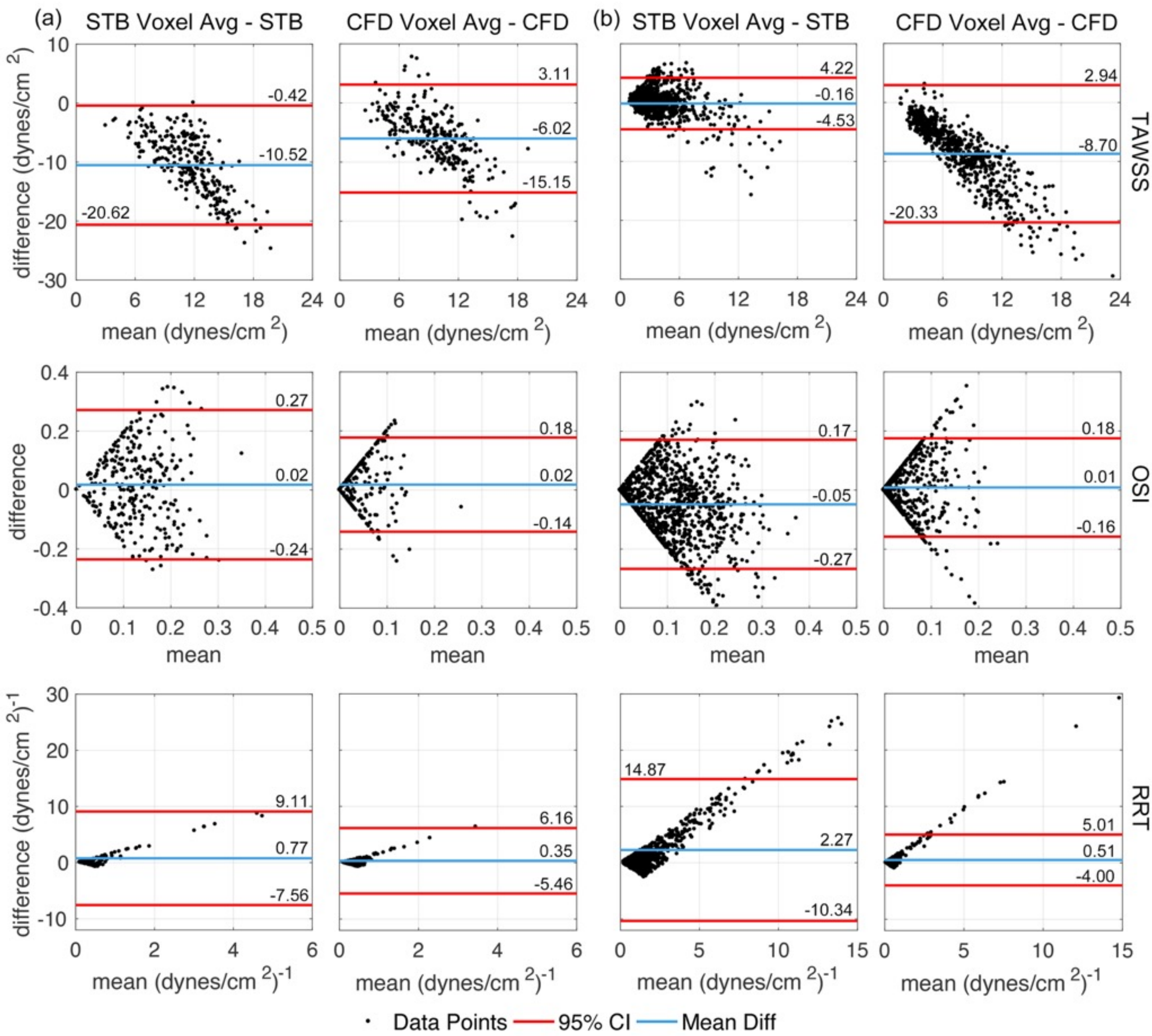

**Figure 7: Bland-Altman analysis of time averaged wall shear stress, oscillatory shear index, and relative residence time, comparing voxel averaged STB to full resolution STB and voxel averaged CFD to full resolution CFD in the (a) basilar tip aneurysm and (b) ICA aneurysm. Mean difference and 95% confidence intervals are indicated.**

With the different behavior of the TAWSS, OSI, and RRT metrics, it is of interest to determine how the metrics perform when comparing the *in vitro* and *in silico* data to the *in vivo* 4D Flow data as well as how the spatial resolution of the *in vitro* and *in silico* datasets affect the comparison. Thus, Bland-Altman analysis was performed to compare these metrics in the aneurysmal sac computed from 4D Flow to those computed from other modalities (Figure 8). In the basilar tip aneurysm (Fig 8a), the mean TAWSS difference was -7.45 and -2.1 dynes/cm$^2$ for the full resolution STB and CFD, respectively. This changed to 1.72 dynes/cm$^2$ for the voxel averaged STB and 2.52 dynes/cm$^2$ for the voxel averaged CFD. The 95% confidence intervals for the voxel averaged data reduced in range by 44% for STB and 31% for CFD as compared to the full resolution intervals. Similar trends were observed in the ICA aneurysm TAWSS (Fig 8b). The mean difference was -0.63 dynes/cm$^2$ for the full resolution STB and -8.92 dynes/cm$^2$ for the full resolution CFD. The 95% confidence interval limits reduced by 51 and 70% for the voxel averaged STB and CFD, respectively, as compared to the full resolution data. Thus, the voxel averaged datasets generally maintained a better match to the 4D Flow TAWSS than the full resolution datasets. In terms of OSI, the mean difference was -0.02 and 0.05 for the full resolution STB and CFD, respectively, in the basilar tip aneurysm. Voxel averaging had a minimal effect, changing the mean difference to 0.01 for CFD and causing no change in STB. Further, the 95% confidence intervals reduced by only 18% for STB and 9% for CFD when voxel averaged. Mean OSI differences of 0.01 for the full resolution STB, 0.07 for the voxel averaged STB, 0.13 for the full resolution CFD, and 0.09 for the voxel averaged CFD were observed in the ICA aneurysm. The interval ranges decreased by 18% when voxel averaged for the STB and increased by 16% when voxel averaged for the CFD. In the basilar tip aneurysm, the mean RRT difference was 0.22 and -0.83 (dynes/cm$^2$)$^{-1}$ for the full resolution and voxel averaged STB, respectively, and 0.18, and -0.02 (dynes/cm$^2$)$^{-1}$ for the full resolution and voxel averaged CFD, respectively. The confidence interval ranges increased 10-fold for the STB when voxel averaged and 3-fold for the CFD when voxel averaged. For the ICA aneurysm, the mean RRT difference was 0.34 (dynes/cm$^2$)$^{-1}$ for the full resolution STB and -0.87 (dynes/cm$^2$)$^{-1}$ for the voxel averaged STB. The mean RRT difference was 1.60 and 0.49 (dynes/cm$^2$)$^{-1}$ for the full resolution and voxel averaged CFD, respectively. The confidence intervals maintained about 4-fold increases in range for the voxel averaged CFD datasets as compared to the full resolution datasets. The RRT confidence interval ranges suggests a better match between the *in vivo* 4D Flow and full resolution datasets, though the mean differences were similar between full and voxel averaged resolutions.

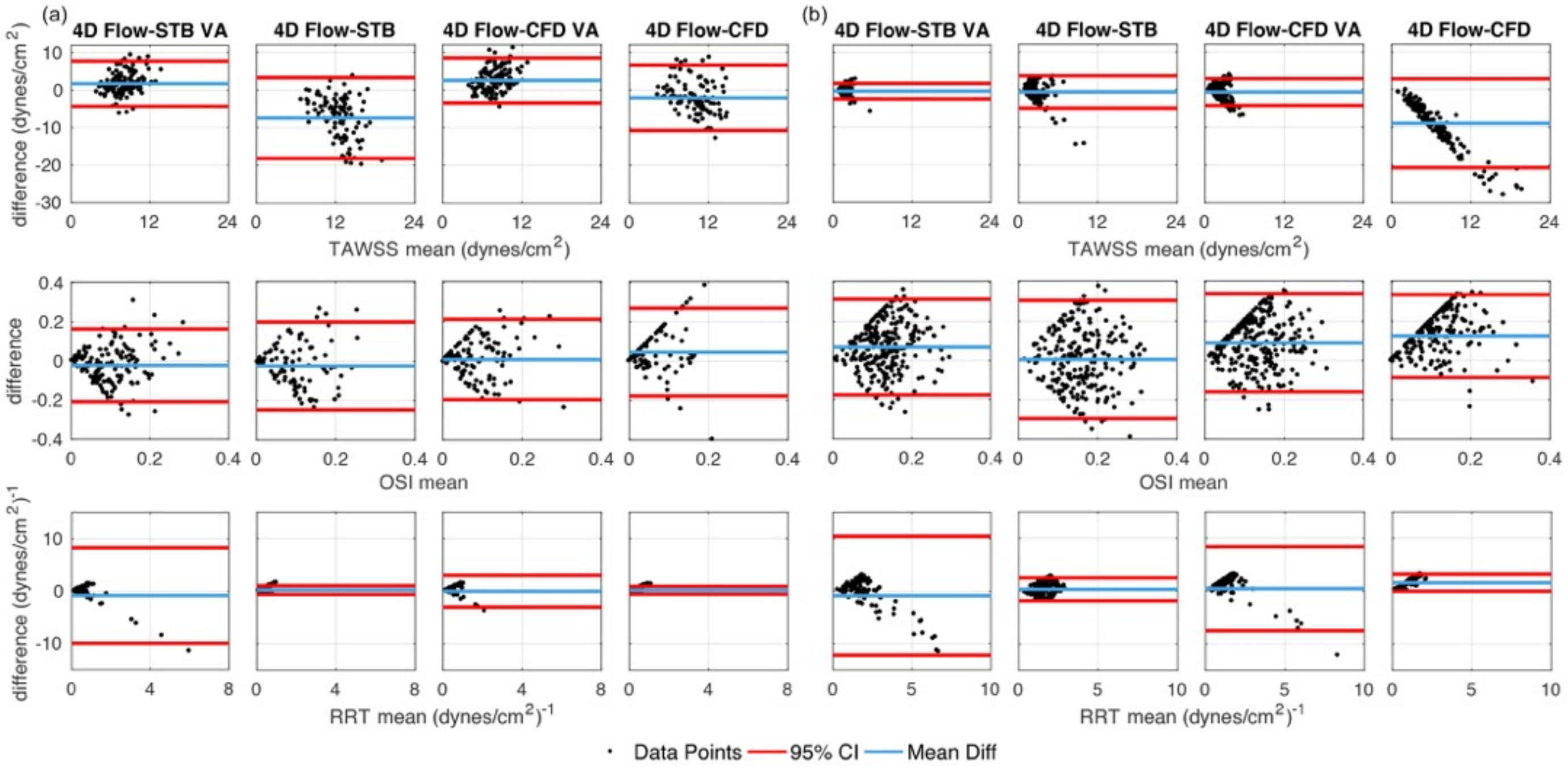

**Figure 8: Bland-Altman analysis of time averaged wall shear stress, oscillatory shear index, and relative residence time, comparing *in vivo* 4D Flow MRI to voxel averaged STB (STB-VA), full resolution STB, voxel averaged CFD (CFD-VA), and full resolution CFD in the (a) basilar tip aneurysm and (b) ICA aneurysm. Mean difference and 95% confidence intervals are indicated.**

# 4 Discussion

The goal of this work was to compare *in vivo* 4D Flow MRI, *in vitro* experimental particle velocimetry, and CFD using two patient-specific cerebral aneurysm models in order to investigate how the limitations, assumptions, and spatial resolutions unique to each modality affected hemodynamic metrics including TAWSS, OSI, and RRT. Experimental PIV studies are significantly less common in this domain than CFD studies, and OSI and RRT have rarely, if ever, been computed using experimental PIV data. However, because experimental studies produce measured rather than simulated velocity fields and typically require less assumptions on the flow, when well controlled, they can often provide a higher fidelity representation of the flow. The use of all three modalities rather than any one or two provided a richer span of the assumptions and limitations that can affect hemodynamic metrics, demonstrating a need for additional comparison studies using volumetric particle velocimetry in this domain. In general, the results of this study demonstrated the limitations and assumptions associated with each modality can have significant effects on hemodynamic metrics. Analysis of TAWSS, OSI, and RRT demonstrated that OSI, a non-dimensional parameter, was most robust to changing modalities (and thus assumptions) and spatial resolutions.

Low WSS regions and their impact on risk of rupture have received much attention in this domain [4,8–11,15]. As seen in Figure 5, all modalities had uniquely different regions of low WSS in the aneurysmal sacs of both geometries, despite the general qualitative agreement of inflow waveforms and velocity pathlines. In this study, WSS was computed using the same methodology, hence the variation cannot be attributed to different calculation biases. Further, Cebral et al. (2011) [14] showed inflow waveform changes can cause variability on the magnitude, but not spatial variation, of hemodynamic metrics. Thus, because the normalized coherent regions of low WSS are of interest, the WSS difference cannot be attributed to variations in inflow rate across modalities. Therefore, this WSS difference was most likely attributable to differences in the assumptions and limitations of each modality. For example, segmentation uncertainties among the *in vitro* and *in silico* data sets, the laminar flow assumptions used in CFD simulations, and the curved extension of the inflow vessel required for experimental model manufacturing for the STB data all could have affected the WSS spatial distributions to some extent. Such assumptions had only small effects on the velocity field and flow patterns but, as shown here, maintained greater variation on the WSS distribution. Thus, the WSS uncertainty resulting from each assumption for each modality must be carefully considered and explicitly studied in future work. Overall, this difference in WSS distribution carries significant clinical and research implications. With the link of low WSS regions to the locations of aneurysm rupture [9,15], the predicted stability of the aneurysm would be different based on the WSS distribution obtained from full resolution CFD versus that obtained from full resolution STB, particularly for the basilar tip aneurysm. Further, studies which investigate hemodynamic variables and analyze which variables have a statistically significant difference between ruptured and unruptured aneurysms, regardless of modality, rely on various flow assumptions which could be largely altering WSS distributions, thus confounding statistically significant results.

The effect of spatial resolution on both velocity and wall shear stress magnitude has been a growing concern. Roloff et al. (2018) [5] demonstrated reductions of velocity magnitude by as much as 10-20% because of voxel averaging. van Ooij et al. (2013) [10] reported WSS values for 4D Flow MRI were lower than that of CFD, but increased if the MRI spatial resolution was increased. Figure 8 agrees with this notion, demonstrating that datasets with similar spatial resolutions have similar TAWSS magnitudes. Specifically, lowering the spatial resolution generally lowers the wall shear stress magnitude. In Figure 7, reduction (bias error) of TAWSS caused by virtual voxel averaging within the same modality was about 6.35 dynes/cm$^2$, corresponding to a roughly 81% error when normalizing by the mean TAWSS across all datasets. Similarly, the average RRT error caused by voxel averaging was 0.98 (dynes/cm$^2$)$^{-1}$, a 91% error when normalizing by the mean RRT across all datasets. However, the average OSI bias error caused by voxel averaging was 0.02, a 26% error when normalized by the mean OSI across all datasets. Further, in Figure 8, when comparing the *in vitro* and *in silico* datasets to the *in vivo* 4D Flow, OSI not only maintained the lowest bias errors (relative to the mean values), but also the most consistent 95% interval bounds across all modalities as well as across full resolution and voxel averaged datasets within the same modality. Thus, OSI was more robust to modality and spatial resolution, and maintained more consistent results than both TAWSS and RRT. OSI yielded no proportional or significant bias errors for all datasets and comparisons. Because OSI is a non-dimensional parameter, its magnitude is less affected by bias error than equivalent dimensional parameters, as bias scale errors in the WSS magnitude would cancel out in the OSI calculation (Equation 2.6). Bias errors are unavoidable across all modalities and subjects and are attributed, at least in part, to the assumptions and spatial resolution of the data as well as calculation methodologies. Thus, OSI and other non-dimensional parameters offer the potential for more consistent risk analysis metrics across a cohort of varying geometries and 4D Flow resolutions than equivalent dimensional parameters. This is a notion that should be explored and expanded in future studies.

There were several limitations of this study. The inlet vessel of the *in vitro* models had to be curved slightly for experimental and manufacturing reasons. This could have had some effect on inflow conditions of the STB experiment, possibly preventing fully developed inflow. Although not ideal, given the natural tortuosity of the cerebral vasculature, the exact *in vivo* inflow conditions would be complex and therefore could not be matched regardless. Further, the resistance valves used to control the outlet flow rates could only adjust the average flow rate in the vessel and not the pulsatility of the flow rate. Thus, some differences in the flow rate range of smaller vessels was observed. Uncertainty quantification methods are widely available for PIV but have yet to be reported for STB and are needed so this data can be used as robust validation test cases for CFD simulations. While virtual voxel averaging was done here, this process does not exactly replicate that of MRI voxel or ensemble averaging. Thus, additional studies are needed to further explore the possible effects of voxel averaging on velocity and post-processing metrics identified in this study using higher fidelity voxel averaging techniques. Bias errors in discrete WSS calculations are well documented in the literature [6,29]. From the WSS algorithm validation done for this study, a consistent bias error was a known issue. Therefore, all WSS, OSI, and RRT values reported in this study can validly be interpreted in a relative sense, but global values of these metrics cannot be extrapolated from this study.

# 5 Conclusion

The goal of this study was to compare *in vitro* volumetric flow velocimetry data with CFD and *in vivo* 4D Flow MRI in order to investigate how the modality and spatial resolution affect TAWSS, OSI, and RRT. The similarity of flow patterns but notable differences in low WSS distributions across modalities demonstrated that more in-depth and comprehensive analysis of specific modeling assumptions and their effect on hemodynamic parameters is needed. Further, non-dimensional parameters, such as OSI, were shown to be more robust and consistent to varying spatial resolutions and modalities, a finding which can improve flow metrics associated with aneurysm risk of rupture in future work. However, given the limitations of this study, additional studies and comparisons are needed to further investigate these findings.

# 6 Acknowledgements

American Heart Association pre-doctoral fellowship (17PRE33670268) to Melissa Brindise, and NIH awards R21 NS106696 and R01 HL115267 (V. Rayz) are gratefully acknowledged.

## 7 Conflict of interest

The authors have no conflicts of interest to report.

## 8 Human and Animal Rights

Patient consent was not required. No human or animal studies were carried out by the authors for this article.